\documentclass[journal]{IEEEtran}
\usepackage{amsmath,amssymb,amsfonts, mathrsfs}
\usepackage[linesnumbered,vlined,ruled]{algorithm2e}
\usepackage{array}
\ifCLASSOPTIONcompsoc
    \usepackage[caption=false, font=normalsize, labelfont=sf, textfont=sf]{subfig}
\else
\usepackage[caption=false, font=footnotesize]{subfig}
\fi
\usepackage{textcomp}

\usepackage{url}
\usepackage{graphicx}
\usepackage{epstopdf}
\usepackage{cite}
\usepackage{booktabs}
\usepackage{multirow}
\usepackage{bigstrut}
\usepackage{color}
\usepackage[
	colorlinks,
	linkcolor=blue,      
	anchorcolor=blue,  
	citecolor=blue]{hyperref}
\graphicspath{{images/}}
\usepackage{subfiles}

\begin{document}

\title{A Computationally Efficient Bi-level Coordination Framework for CAVs at Unsignalized Intersections}




\author{Jiping~Luo,~
        Tingting~Zhang,~\IEEEmembership{Member,~IEEE,}
        and~Qinyu~Zhang,~\IEEEmembership{Senior~Member,~IEEE}

%
}

\maketitle

\begin{abstract}
	In this paper, we investigate cooperative vehicle coordination for connected and automated vehicles (CAVs) at unsignalized intersections. To support high traffic throughput while reducing computational complexity, we present a novel collision region model and decompose the optimal coordination problem into two sub-problems: \textit{centralized} priority scheduling and \textit{distributed} trajectory planning. Then, we propose a bi-level coordination framework which includes: (i) a Monte Carlo Tree Search (MCTS)-based high-level priority scheduler aims to find high-quality passing orders to maximize traffic throughput, and (ii) a priority-based low-level trajectory planner that generates optimal collision-free control inputs. Simulation results demonstrate that our bi-level strategy achieves near-optimal coordination performance, comparable to state-of-the-art centralized strategies, and significantly outperform the traffic signal control systems in terms of traffic throughput. Moreover, our approach exhibits good scalability, with computational complexity scaling linearly with the number of vehicles. Video demonstrations can be found online at \url{https://youtu.be/WYAKFMNnQfs}.
\end{abstract}
\begin{IEEEkeywords}
connected and automated vehicles (CAVs), intersection management, cooperative vehicle control.
\end{IEEEkeywords}

\section{INTRODUCTION}
\IEEEPARstart{B}{enefiting} from advances in autonomous driving and vehicular communications\cite{survey_5g_nr,survey_vehicular_transportation, survey_autonomous_driving}, connected and automated vehicles (CAVs) have emerged as a key enabler of intelligent transportation systems (ITS)\cite{survey_cooperative_architecture, survey_intersection_management, survey_khayatian,survey_zhongzijia}. By allowing vehicles to share information and make decisions cooperatively via vehicle-to-everything (V2X) communication links, cooperative vehicle coordination will be a promising alternative to traditional traffic signal control systems, and is envisioned to make everyday traveling safer, greener, and more efficient\cite{survey_chenlei, survey_cav_intersection_highway, survey_hult}. 

Over the last two decades, \textit{centralized} coordination strategies have been widely studied to enable optimal maneuvers for CAVs at unsignalized intersections. In this paradigm, a coordination center, usually fulfilled at the road side unit (RSU), collects the driving status (e.g., velocities, accelerations, positions, routes, etc) of all involved vehicles and plans safe controls for them\cite{luo_iccc,chunsheng}. Generally, the central goal is to maximize traffic throughput subjecting to dynamics/kinematics, actuator limits, and collision avoidance constraints (including rear-end and lateral safety). Since all vehicles are considered as one system, centralized strategies could generate globally optimal trajectories that are collision-free and deadlock-free. However, centralized strategies confront challenges due to the contradiction between stringent real-time actuation requirements and the high computational complexity required for optimizations \cite{complecity_np_hard, hult_complexity}.

\begin{figure*}[t]
	\centering
	\subfloat[collision set]{%
		\includegraphics[scale=0.2]{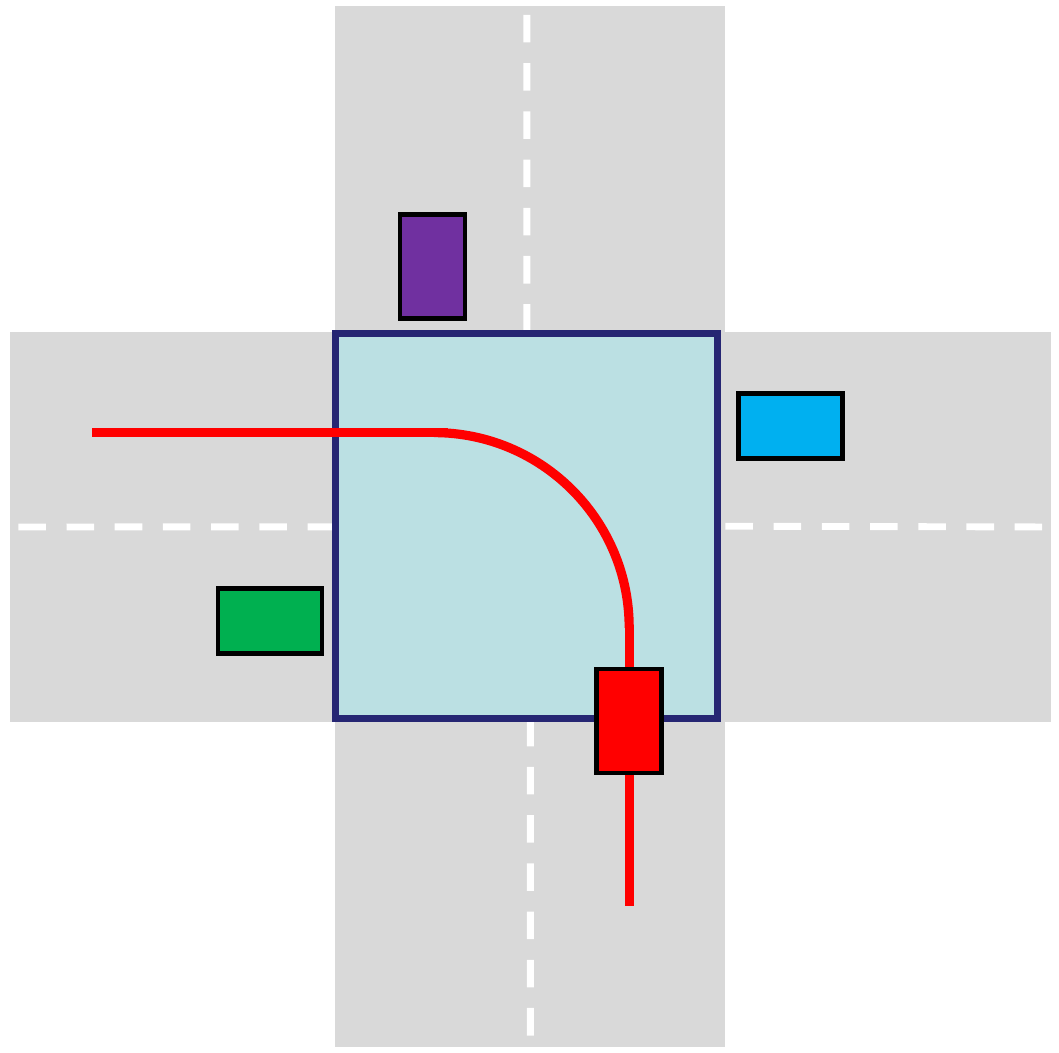}}
	\hspace{0.8em}
	\subfloat[cross-collision points]{%
		\includegraphics[scale=0.2]{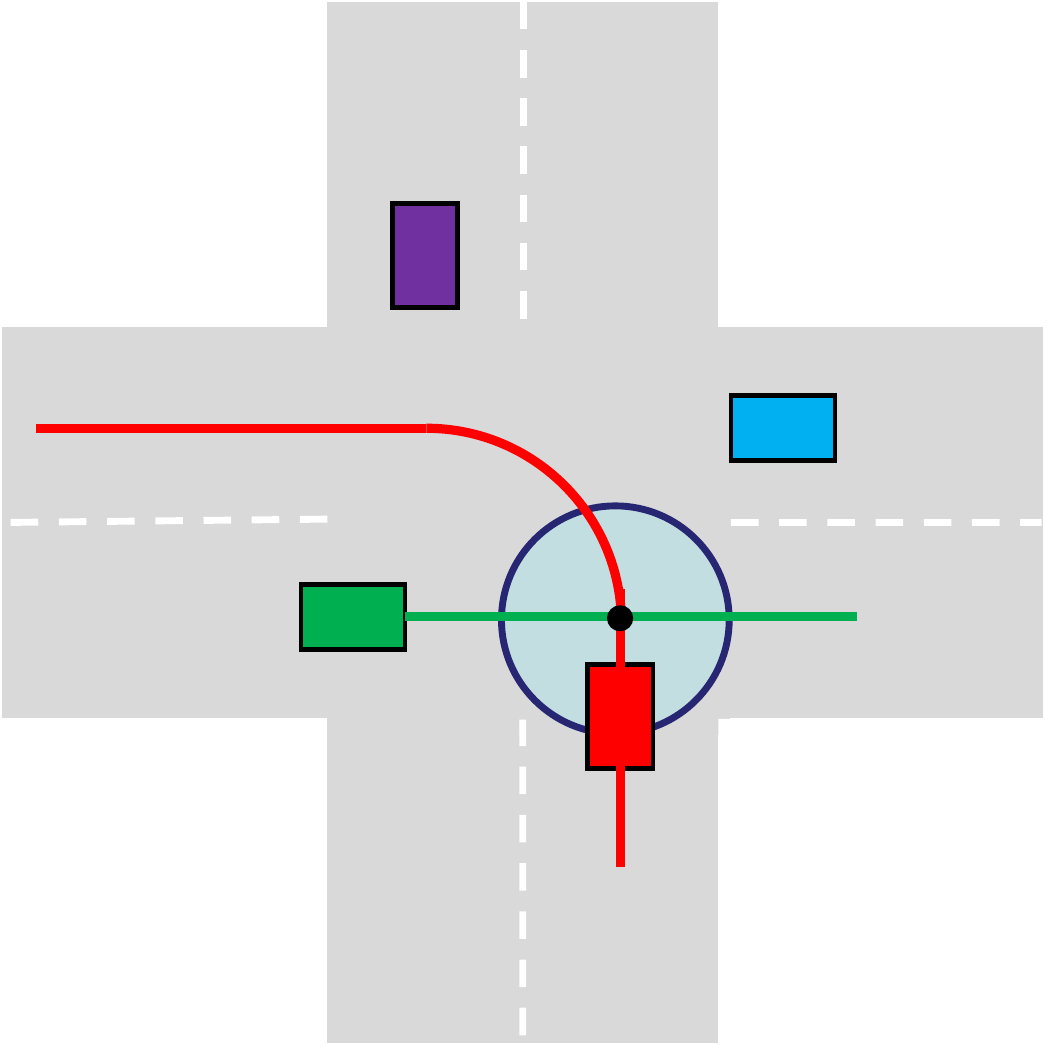}}
	\hspace{0.8em}
	\subfloat[rectangular sub-zones]{%
		\includegraphics[scale=0.2]{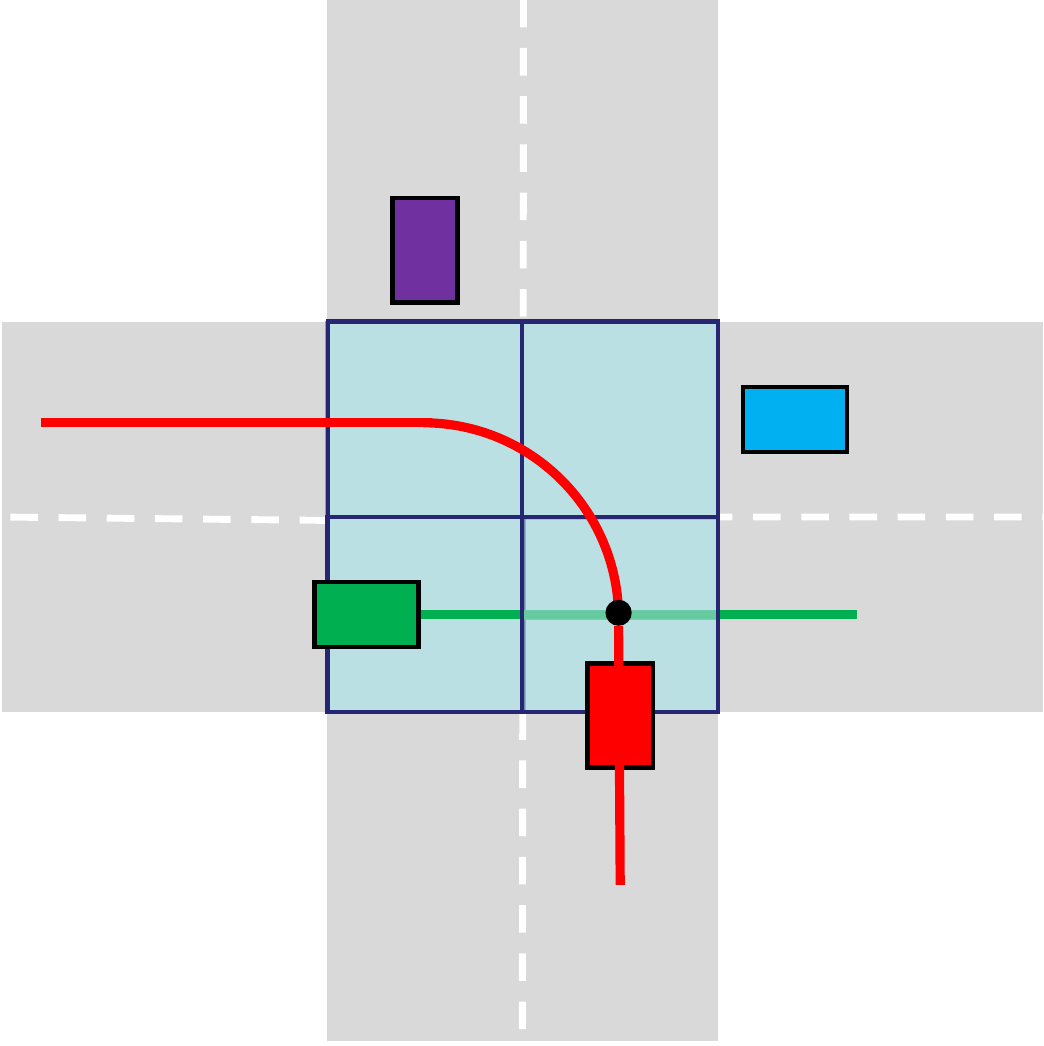}}
	\hspace{0.8em}
	\subfloat[space-time resource searching]{%
		\includegraphics[scale=0.2]{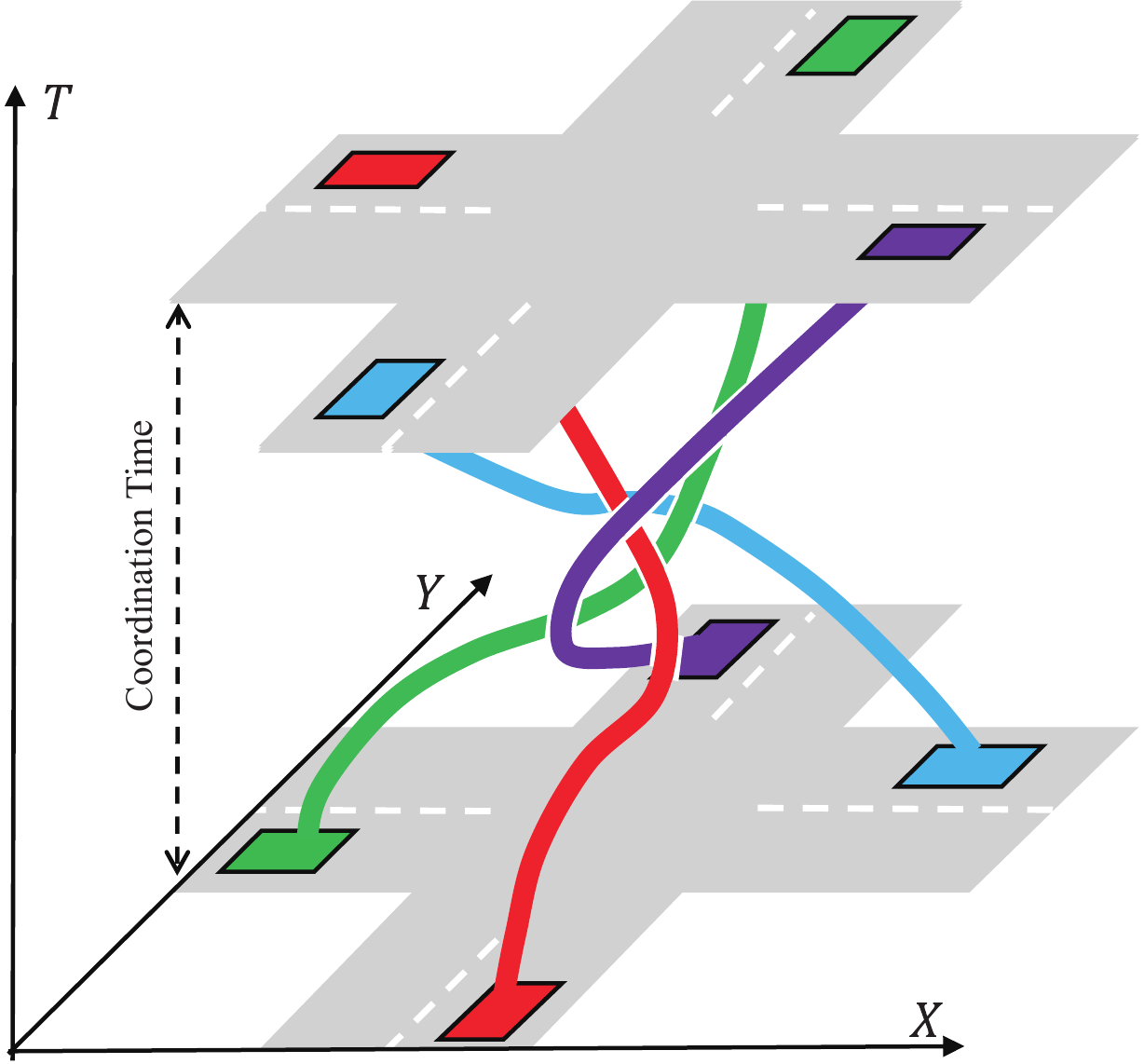}}
	\caption{Description of collision region models for a single-lane intersection.}
	\label{fig:strategies}
\end{figure*}

At intersections, traffic throughput and safety are closely intertwined and need to be jointly considered. One classical formulation for safety is the \textit{collision set (CS)} model\cite{lee_cs, hult_cs, wangmengqi_cs}, as shown in Fig.~\ref{fig:strategies}(a). In this model, the whole intersection area was defined as a collision set (i.e., blue area) where any two vehicles from conflicting routes cannot exist in the area simultaneously. By indicating vehicle priorities using \textit{auxiliary binary variables} \cite{hult_cs,wangmengqi_cs,liuchanghao}, the optimal coordination problem can be formulated as a Mixed-integer Quadratic Programming (MIQP) problems. Although optimal passing orders and vehicle trajectories can be obtained simultaneously, such centralized CS based approaches are still far from satisfactory due to the following two limitations. Firstly, excessive safety redundancy is reserved for only one vehicle at a time, thus reducing the utilization of intersection spatial resources and traffic efficiency. Secondly, MIQPs are computationally prohibitive or even intractable since the time complexity increases exponentially with the number of vehicles. To approach these gaps, several collision region models and low-complexity coordination algorithms have been proposed in recent years.

To address the underutilization issue of the intersection spatial resources, several collision region models have been proposed. The authors in \cite{CCP, liuchanghao, moyangan} defined multiple \textit{cross-collision points (CCPs)} inside the intersection area. The circle area centering the CCP show an approximate area where any two conflicting vehicles are not allowed to enter at the same time, as shown in Fig.~\ref{fig:strategies}(b). Similarly, the intersection area can also be divided into multiple \textit{rectangular sub-zones} according to the combination of each pair of orthogonal lanes\cite{subzones, hult_complexity}, as shown in Fig.~\ref{fig:strategies}(c). However, all the aforementioned collision region models regard vehicles as mass points and provide excessive safety redundancy compared to the vehicle geometry. More recently, a three-dimensional collision region model called \textit{space-time resource searching (STRS)} was proposed in \cite{STRS}. This model stretches the intersection capacity to its limit by allocating non-overlapping space-time resources to vehicles in the XYT domains, as shown in Fig.~\ref{fig:strategies}(d). Although these models demonstrate remarkable coordination efficiency advantages compared to the CS model, the corresponding centralized coordination strategies often suffer from overly long problem-solving times due to the need to avoid collisions at each sub-region.

To address the key computational challenges in real-world
implementations, in our previous contributions\cite{luo_tits}, we proposed a deep reinforcement learning (DRL)-based strategy to generate cooperative trajectories for a batch of vehicles in real-time. Although near-optimal traffic efficiency can be achieved in sub-static coordination scenarios, the introduction of vehicle grouping and virtual vehicle padding methods may hinder this strategy for high-traffic loads. A number of research efforts have developed \textit{distributed}\footnote{We follow the definition from \cite{definition}, where "distributed control" refers to vehicles exchanging local states and decisions to collaboratively accomplish a task, whereas in "decentralized control", vehicles independently determine their actions without exchanging information.} strategies to reduce computation time. In these methods, vehicles are firstly prioritized according to some basic rules, such as first-in-first-out (FIFO)\cite{fifo}, distance to intersection\cite{distance_to_intersection}, time to react\cite{time_to_react_2, time_to_react_3}, etc. Then each vehicle solves its part of optimization problem sequentially to avoid collisions with high-priority neighbors. In this way, the optimal coordination problem is decomposed into a sequence of low-complexity trajectory planning sub-problems and the exponential complexity can be avoided. However, due to the lack of global information, such rule-based priority assignment methods may result in inefficient coordination performances.

To achieve a good trade-off between complexity and optimality, a number of research efforts have recently developed bi-level coordination strategies. The main idea behind these strategies is to use a \textit{centralized} layer to assign collision-free time slots to vehicles, along with a \textit{decentralized} layer that follows the predetermined vehicle motion patterns \cite{malikopoulos2021optimal,chen2021graph,cong2022virtual}. For instance, the authors in \cite{malikopoulos2021optimal} derived fuel-efficient piece-wise analytic speed profiles, based on which formed an efficient centralized problem to resolve collisions and improve throughput. Another method is the virtual platoon \cite{chen2021graph,cong2022virtual}, which simplifies trajectory planning by projecting vehicles from different lanes onto a single virtual lane and constructing virtual platoons on this lane to maintain typical platooning behavior. Similar methods are employed in \cite{faris2022optimization,deng2023cooperative}, which leverage vehicle platooning schemes for dealing with mixed traffic intersection coordination. However, these strategies require predetermined rules about vehicle motion patterns, thereby limiting degrees of freedom available for vehicle maneuvering and reducing coordination efficiency.






In this paper, we aim to develop a high degree of freedom bi-level coordination framework that combines the strengths of centralized and distributed strategies. Our aim is to improve traffic throughput while reducing computational complexity. The main contributions are summarized as follows.
\begin{itemize}
	\item We propose a novel bi-level coordination framework that decomposes the optimal coordination problem into two sub-problems: \textit{centralized} priority scheduling and \textit{distributed} trajectory planning. The centralized coordinator only assigns priorities to vehicles, whereas each vehicle thereafter plan and share its own trajectory in sequence to avoids collisions with high-priority vehicles. 
	\item To fully utilize the spatial resources and improve coordination efficiency, we introduce a novel collision region model that leverages collision-checking algorithms to pre-compute collision boundaries for any two conflicting paths. Additionally, we propose a low-complexity distributed trajectory planner that enables vehicles to cross intersections safely and efficiently. 
	\item To tackle the challenging combinatorial priority scheduling problem, we propose a Monte Carlo tree search (MCTS)-based high-level planner that efficiently identifies near-optimal priority sequences within a limited computational budget. 
	\item Our proposed bi-level strategy has been evaluated through simulations, demonstrating comparable coordination performance to state-of-the-art centralized strategies while significantly reducing computational complexity.
\end{itemize}

The remained of this paper is organized as follows. The system model and the bi-level coordination framework are presented in Section \ref{sec:system_model}. The low-level distributed trajectory planner and the high-level priority scheduler are presented in Section \ref{sec:low_level_planner} and Section \ref{sec:high_level_planner}, respectively. Section \ref{sec:results} presents our simulation results to confirm the capability and scalability of the proposed strategy. Finally, we conclude this paper and discuss future directions in Section \ref{sec:conclusion}.

\section{System Model}\label{sec:system_model}

\subsection{Intersection Model}
\begin{figure}[h]
	\centering
	\includegraphics[width=0.9\linewidth]{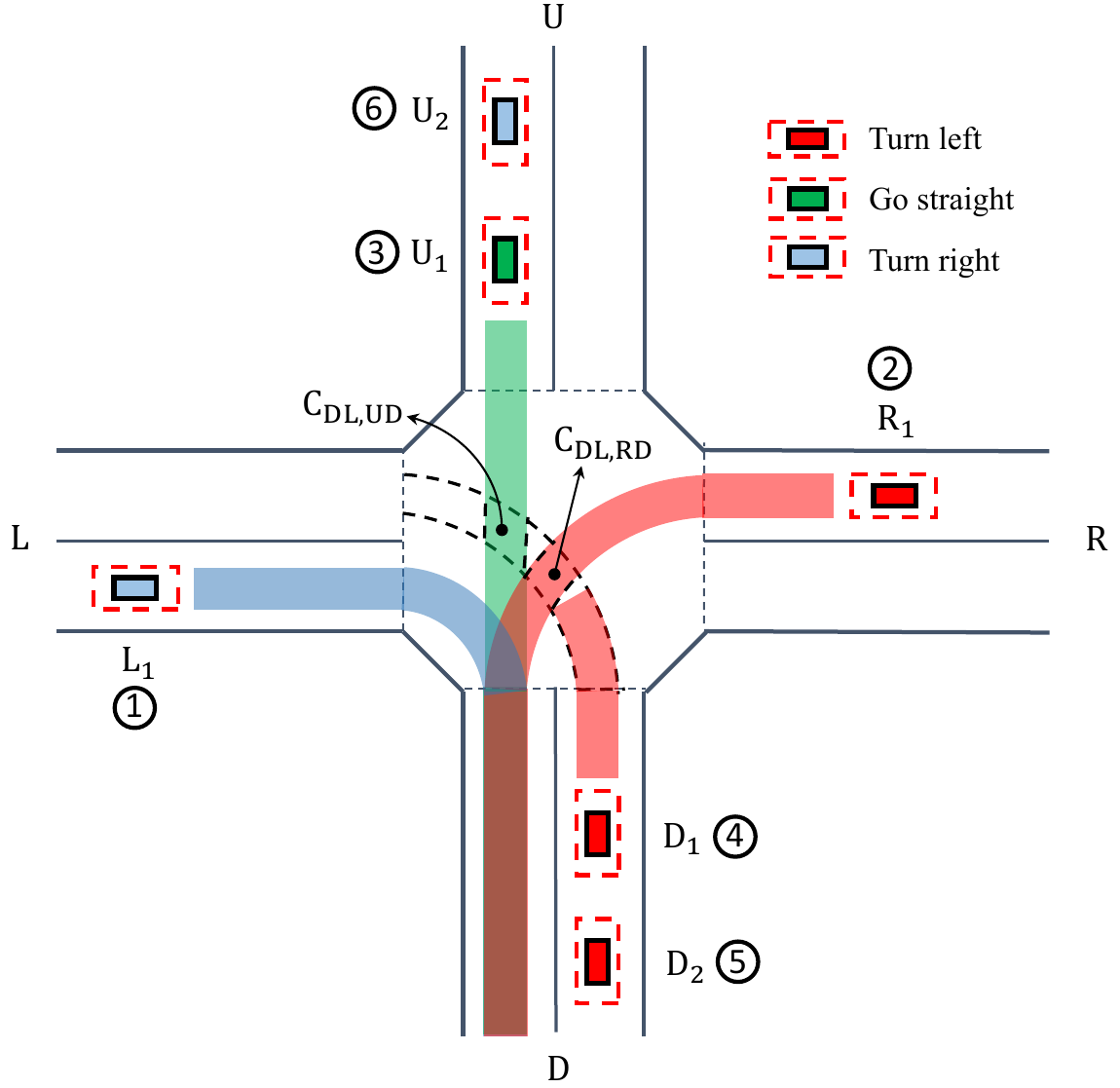}
	\caption{A typical signal-free intersection scenario.}
	\label{fig:systemmodel}
\end{figure}

In this paper, we consider a typical single-lane intersection scenario which consists of 4 roads, i.e., down road, right road, up road, and left road, as shown in Fig.~\ref{fig:systemmodel}. On each road, there are three possible paths that vehicles can follow: go straight, turn left and turn right, and there is a total of 12 different paths within the intersection area. The central area where multiple paths cross and merge is called the conflict area (CA). A centralized coordinator is deployed at the RSU to store road structure information (e.g., intersection's geometric parameters, pre-defined paths, etc), collect vehicles' status information (e.g., positions, velocities, accelerations, and routes, etc), and give instructions to vehicles to cross the intersection safely and efficiently. $\mathrm{C}_{\mathrm{DL,RD}}$ represents the collision region of two conflicting paths: ego vehicle's path $\mathrm{D}\rightarrow\mathrm{L}$ and high-priority vehicle's  $\mathrm{R}\rightarrow\mathrm{D}$. The calculation of all collision regions will be discussed in the next subsection. 

Without loss of generality, we assume that vehicles must follow one of the 12 pre-defined paths according to their driving directions and only the longitudinal movements of vehicles are controlled. Hence, the dynamics of each vehicle $i \in \mathcal{N}$ can be modeled as a double integrator, i.e.,
\begin{align}
	\dot{s}_i (t) = v_i (t) \label{eq:kinematic_1}\\
	\dot{v}_i (t) = u_i (t) \label{eq:kinematic_2}
\end{align}
where $s_i(t)$ denotes the position along its path, $v_i(t)$ and $ u_i (t)$ are the longitudinal speed and acceleration. Considering the practical constraints in the vehicle dynamics, we do have:
\begin{align}
	[s_i(0), v_i(0&), u_i(0)] = [s_i^0, v_i^0, u_i^0]\\
	0 &\leq v_i(t) \leq v_\mathrm{max} \label{eq:kinematic_3} \\
	u_\mathrm{min} &\leq u_i(t) \leq u_\mathrm{max} \label{eq:kinematic_4} 
\end{align} 
here $[s_i^0, v_i^0, u_i^0]$ is the initial state of vehicle $i$, which should be collected by the coordination center at the beginning. 

\subsection{Collision Region Model}\label{subsection:collision_region_model}

\begin{figure}[h]
	\centering
	\includegraphics[width=\linewidth]{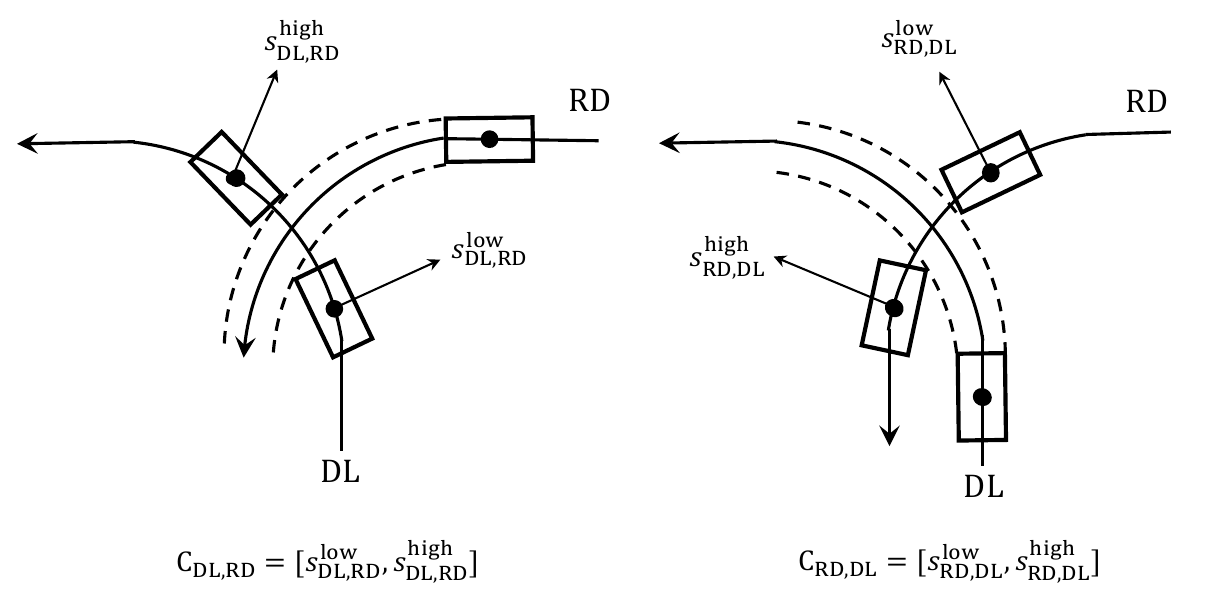}
	\caption{An illustration of the proposed collision region model.}
	\label{fig:collisiontable}
\end{figure}

Collision region modeling is essential for coordination safety and efficiency. To fully utilize the spatial resources of the intersection area, we propose a novel collision region model by leveraging collision-checking algorithms. Fig.~\ref{fig:collisiontable} demonstrates the calculation of collision region between two conflicting routes $\mathrm{D}\rightarrow\mathrm{L}$ and $\mathrm{R}\rightarrow\mathrm{D}$. It is worth noting that $\mathrm{C}_{\mathrm{DL,RD}}$ and $\mathrm{C}_{\mathrm{RD,DL}}$ are two different collision regions since the ego vehicle is on different paths. Take $\mathrm{C}_{\mathrm{DL,RD}}$ as an example, we can locate the ego vehicle at a given point $s_\mathrm{ego}$ on path DL, and let conflicting vehicle travel at a constant speed along path RD. We then use collision-checking algorithms such as Separating Axis Theorem (SAT)\cite{SAT} to check if a collision occurs. By iteratively locating the ego vehicle from one side to another along path DL, we can get the exact collision boundary $s_{\mathrm{DL,RD}}^\mathrm{in}$ that the ego vehicle cannot enter before the conflicting vehicle passes through. In this case, the ego vehicle can enter the collision region $\mathrm{C}_{\mathrm{DL,RD}} = [s_{\mathrm{DL,RD}}^\mathrm{in}, s_{\mathrm{DL,RD}}^\mathrm{out}]$ if and only if the high-priority vehicle pass its collision region $\mathrm{C}_{\mathrm{RD,DL}} = [s_{\mathrm{RD, DL}}^\mathrm{in}, s_{\mathrm{RD,DL}}^\mathrm{out}]$. This calculation process is time-consuming but can be pre-computed since road structures and paths are fixed. The collision table $\mathcal{T}_\mathrm{collision}$ consisting of all collision regions can be obtained from Algorithm \ref{alg:collisiontable}. Herein, each pair of paths intersect with each other at most once and we do not consider the self-collision of two identical paths.

\begin{algorithm}
	\caption{Offline Collision Table Generation.}
	\label{alg:collisiontable}
	\KwIn{pre-defined paths $\mathcal{P}_\mathrm{route}$}
	\KwOut{collision table $\mathcal{T}_\mathrm{collision}$}
	\For{$p_{\mathrm{ego}} \in \mathcal{P}_\mathrm{route}$}{
		\For{$p_\mathrm{conflict} \in \mathcal{P}_\mathrm{route}, \, p_\mathrm{ego} \neq p_\mathrm{conflict}$}{
			$a \leftarrow s_\mathrm{ego}^{\mathrm{init}}, b \leftarrow s_\mathrm{ego}^{\mathrm{end}}$ \;
			\While{$|a - b| \geq eps$}{
				Locate ego vehicle at $c \leftarrow (a + b) / 2$ \;
				Let conflicting vehicle travel along $p_\mathrm{conflict}$\;
				Check if collision occurs using SAT\;
				\lIf{collide}{$b \leftarrow c$}
				\lElse{$a \leftarrow c$}
			}
			$s_\mathrm{ego}^\mathrm{in} = a$\;
			Set $a \leftarrow s_\mathrm{ego}^{\mathrm{end}}, b \leftarrow s_\mathrm{ego}^\mathrm{in}$, repeat line 4-9\;
			$s_\mathrm{ego}^\mathrm{out} = a$\;
			$\mathcal{T}_\mathrm{collision}[p_\mathrm{ego}][p_\mathrm{conflict}] = [s_\mathrm{ego}^\mathrm{in}, s_\mathrm{ego}^\mathrm{out}]$\;
		}
	}
\end{algorithm}

\subsection{Centralized Coordination Strategies}\label{sec:centralizedcs}
In this section, we first introduce a classical centralized coordination strategy based on the CS model\cite{hult_cs}, as shown in Fig.~\ref{fig:strategies}(a). Considering that any two conflicting vehicles must cross the collision set one after another, the lateral collision-free constraints can be written as:          
\begin{align}
	[t_i(s_i^\mathrm{in}), t_i(s_i^\mathrm{out})] \cap [t_j(s_j^\mathrm{in}), t_j(s_j^\mathrm{out})] = \emptyset, \forall j \in \mathcal{N}_\mathrm{con}^i \label{eq:cslateralconstraint_1}
\end{align}
where $\mathcal{N}_\mathrm{con}^i$ denotes the set of all conflicting vehicles of vehicle $i$, $s^\mathrm{in}$ and $s^\mathrm{out}$ are the entry and exit point of the collision set, respectively. For a given initial configuration, the exact time that a vehicle occupies the collision set is determined by the passing order and trajectory planning results. However, the optimal ordering for a given initial configuration can only be found by evaluating all feasible passing order alternatives in a structured manner. Consequently, constraint \eqref{eq:cslateralconstraint_1} is essentially NP-hard, and it is necessary to seek low-complexity relaxations. One common approach is introducing auxiliary binary variables, i.e., $\delta_i(t), \gamma_i(t) \in \{0, 1\}$, to indicate the priority of each vehicle. For instance, $\delta_i(t) = 1, \gamma_i(t) = 1$ indicates that vehicle $i$ is inside the CA, while $\delta_i(t) = 0, \gamma_i(t) = 1$ or $\delta_i(t) = 1, \gamma_i(t) = 0$ means that vehicle $i$ has not yet reached the CA or has already passed the CA. $\delta_i(t) = 0, \gamma_i(t) = 0$ is invalid state. Thus, we can replace constraint \eqref{eq:cslateralconstraint_1} using the following linear inequalities:
\begin{align}
	s_i^\mathrm{out} - \gamma_i (t) M \leq s_i(t) \leq s_i^\mathrm{in} + \delta_i(t)M \label{eq:cslateral_1}\\
	\delta_i(t) + \gamma_i(t) + \delta_j(t) + \gamma_j(t) \leq 3  \label{eq:cslateral_2}
\end{align}
where $M$ is a sufficiently large number. 

To guarantee no rear-end collision occurs between vehicles on the same lane, we impose the following rear-end safety constraint:
\begin{align}
	s_j(t) - s_i(t) \geq L_\mathrm{safe} \label{eq:csrear_end}
\end{align}
where $j \in \mathcal{N}_\mathrm{pre}^i$ is the preceding vehicles in front of vehicle $i$, $L_\mathrm{safe}$ is the longitudinal safety gap.

Thus, the centralized-CS problem with the aim of maximizing traffic throughput can be formulated as:
\begin{align*}
	\mathscr{P}_\mathrm{centralized-CS}: \, \, &\min_{u_i(t), \delta_i(t), \gamma_i(t)} \,\, \sum_{i=1}^{N} \sum_{t=0}^{T-1} (v_i(t) - v_\mathrm{max})^2 \\
	\mathrm{s.t.}\,\,\, 
	&\,\,\eqref{eq:kinematic_1} - \eqref{eq:kinematic_4},\,\,\, \forall i \in \mathcal{N}\\ 
	&\,\,\eqref{eq:cslateral_1}  - \eqref{eq:cslateral_2},\,\,\,  \forall i \in \mathcal{N}, j \in \mathcal{N}_\mathrm{con}^i\\
	&\,\,\eqref{eq:csrear_end},\,\,\, \forall i \in \mathcal{N}, j \in \mathcal{N}_\mathrm{pre}^i
\end{align*}

Similarly, we can formulated a centralized optimal coordination problem (OCP) based on our collision region model described in Section \ref{subsection:collision_region_model}. The lateral collision avoidance constraints are as follows,
	\begin{align}
		s_{i,c}^\mathrm{out} - \gamma_{i,c}(t) M \leq s_i(t) \leq s_{i,c}^\mathrm{in} + \delta_{i,c}(t)M \label{eq:ccplateral_1}\\
		\delta_{i,c}(t) + \gamma_{i,c}(t) + \delta_{j,c}(t) + \gamma_{j,c}(t) \leq 3  \label{eq:ccplateral_2}
	\end{align}
	where $c \in \mathcal{C}_\mathrm{con}^{i,j}$ is the corresponding collision region between vehicle $i$ and vehicle $j$, $s_{i,c}^\mathrm{in}$ and $s_{i,c}^\mathrm{out}$ are the entry and exit point of collision region $c$. As a result, the centralized optimal coordination problem (OCP) can be formulated as follows,
	\begin{align*}
		\mathscr{P}_\mathrm{centralized-OCP}:& \, \, \min_{u_i(t), \delta_{i,c}(t), \gamma_{i,c}(t)} \,\, \sum_{i=1}^{N} \sum_{t=0}^{T-1} (v_i(t) - v_\mathrm{max})^2 \\
		\mathrm{s.t.}\,\,\, 
		&\,\,\eqref{eq:kinematic_1} - \eqref{eq:kinematic_4},\,\,\, \forall i \in \mathcal{N}\\ 
		&\,\,\eqref{eq:csrear_end},\,\,\, \forall i \in \mathcal{N}, j \in \mathcal{N}_\mathrm{pre}^i \\
		&\,\,\eqref{eq:ccplateral_1}  - \eqref{eq:ccplateral_2},\,\,\,  \forall i \in \mathcal{N}, j \in \mathcal{N}_\mathrm{con}^i, c \in \mathcal{C}_\mathrm{con}^{i,j}
\end{align*}
Problems $\mathscr{P}_\mathrm{centralized-CS}$ and $\mathscr{P}_\mathrm{centralized-OCP}$ can be cast into Mixed-integer Quadratic Programming (MIQP) problems, which can return optimal inputs $u$ and priority sequence $\gamma, \delta$ simultaneously. However, MIQPs are generally time-consuming and the problem size may grow exponentially with the number of vehicles.

%

\subsection{Priority-based Bi-level Coordination Framework}
\begin{figure}[t]
	\centering
	\includegraphics[width=\linewidth]{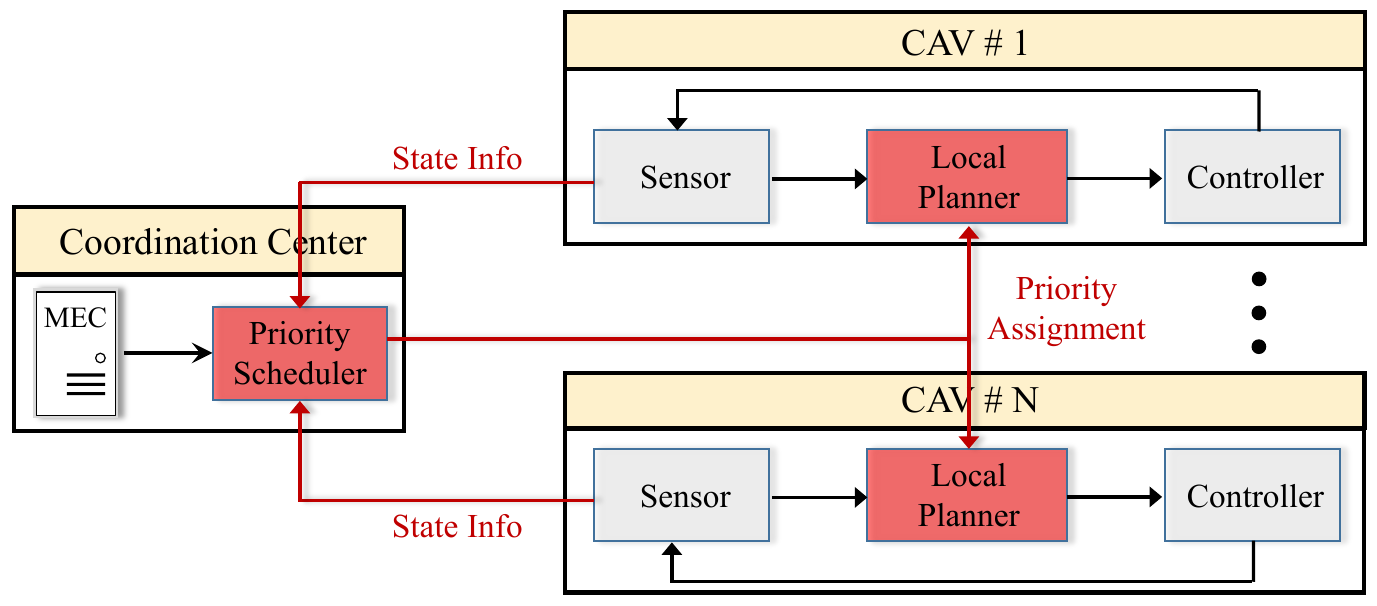}
	\caption{The proposed bi-level coordination framework.}
	\label{fig:bilevelcoordinator}
\end{figure}

Although the aforementioned centralized strategies are able to find global optima by jointly optimizing cooperative trajectories and priority sequence, they scale poorly and are somehow not directly applicable in industrial practice. In this work, we propose an efficient bi-level coordination framework, as shown in Fig.~\ref{fig:bilevelcoordinator}. 

The centralized node (high-level planner) is deployed to collect system-wide state information of all involved vehicles. Based on the initial configuration, the high-level planner only solves the priority scheduling problem and then assigns priority to vehicles. Note that adjacent non-conflicting vehicles may be assigned with the same priority, while conflicting vehicles must be sequentially arranged.
The low-level planner of each vehicle thereafter solves its local trajectory planning problem in sequence to avoid collisions with high-priority vehicles. Then, the optimized trajectory is transmitted to low-priority neighbors for their planning problems.

\begin{figure*}[t]
	\centering
	\begin{minipage}[t]{0.4\textwidth}
		\includegraphics[width=0.9\linewidth]{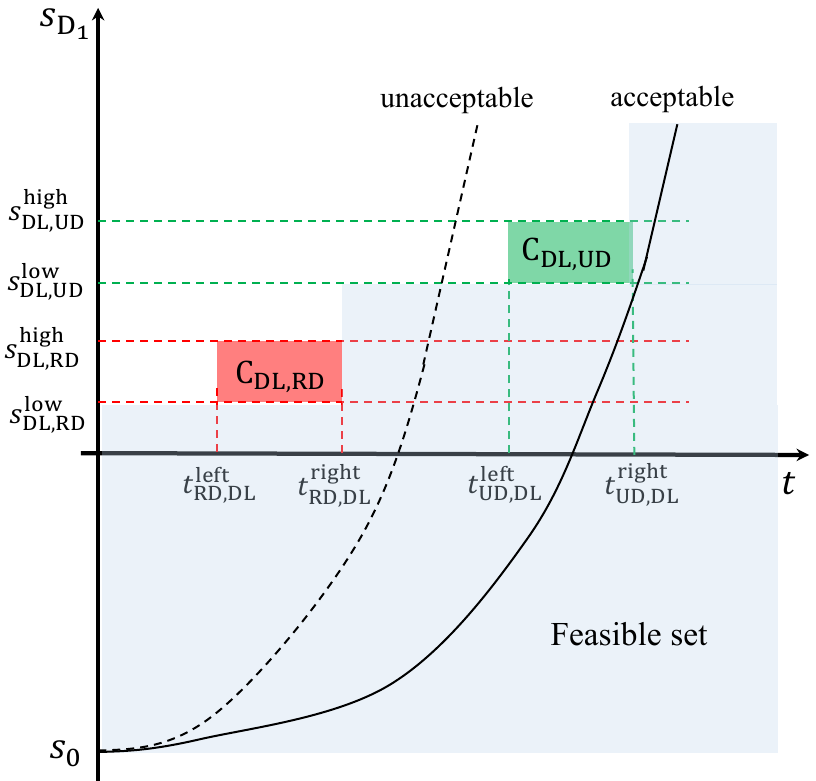}
		\caption{S-T graph based low-complexity trajectory planner.}
		\label{fig:stplanner}
	\end{minipage}
	\hfill
	\begin{minipage}[t]{0.58\textwidth}
		\centering
		\includegraphics[width=1\linewidth]{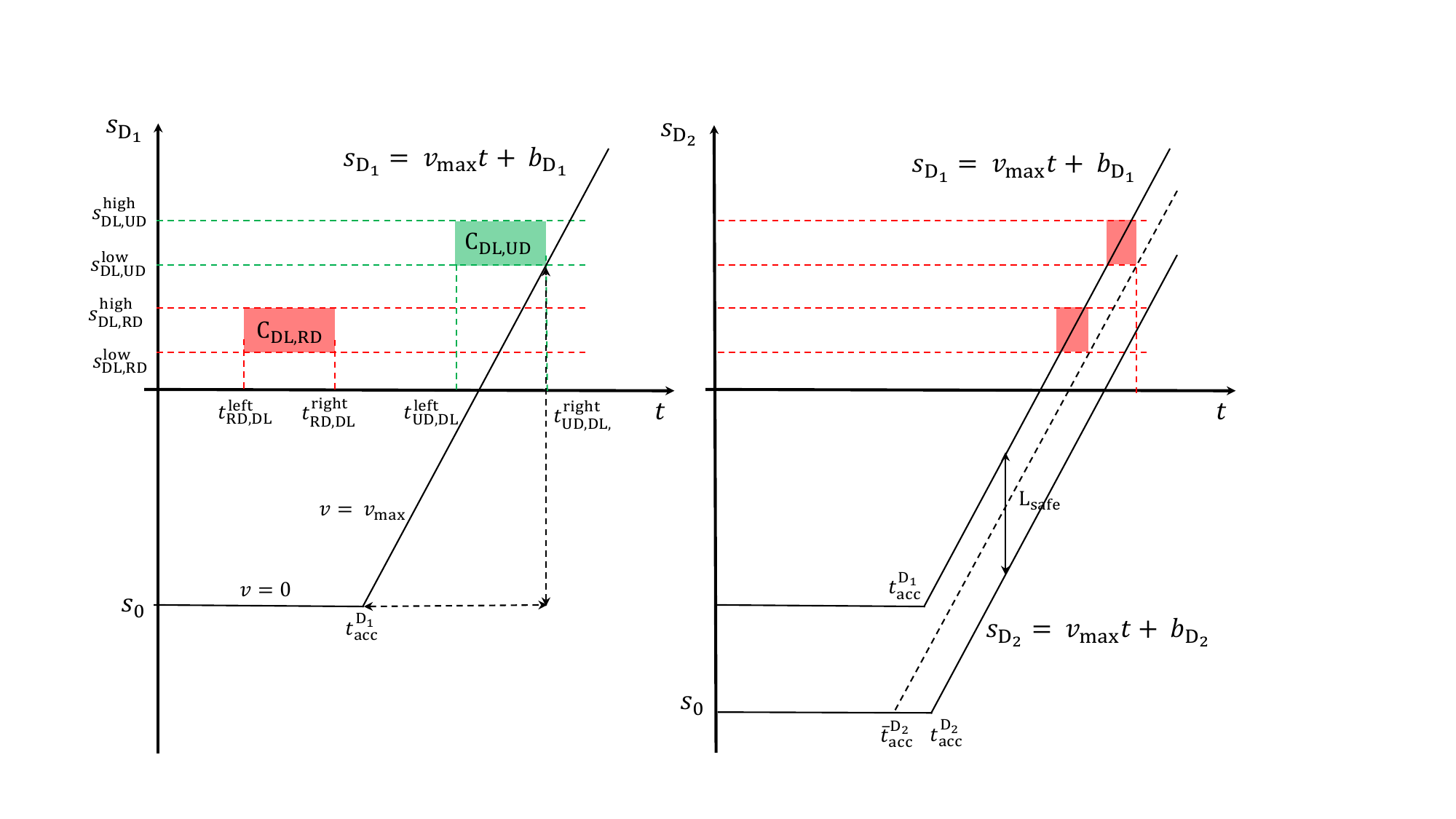}
		\caption{Calculation of the simplified speed profiles.}
		\label{fig:simplifiedstplanner}
	\end{minipage}
\end{figure*}

The key issue to be addressed is decomposing the optimal coordination problem into the priority scheduling and distributed trajectory planning. The quality (i.e., traffic efficiency) of a given priority sequence can only be evaluated after all vehicles have planned their trajectories, which means priority scheduling is closely intertwined with trajectory planning. Rule-base approaches\cite{fifo,distance_to_intersection,time_to_react_2,time_to_react_3} greedily find a feasible priority sequence without evaluating the quality of the solution which may, however, greatly sacrifice traffic efficiency. Enumeration-based approaches\cite{STRS} find the globally optimal solution by evaluating all possible alternatives. However, they are not applicable in practice since the priority scheduling itself is a combinatorial optimization problem and iterative trajectory planning is also time-consuming. 

In the following two sections, we will proposed a promising scheme to address these challenges. We start our exposition with the low-level trajectory planner in Section \ref{sec:low_level_planner} and then we discuss the high-level priority scheduler in Section \ref{sec:high_level_planner}.

\section{Low-level Distributed Trajectory Planner}\label{sec:low_level_planner}
In this section, we consider that the solution of the high-level problem is given. We now focus on the low-level distributed trajectory planning problem. Our proposed low-level planner follows the principle that each vehicle needs to give way to high-priority vehicles. Take the coordination scenario described in Fig.~\ref{fig:systemmodel} as an example. For a given priority sequence $(\mathrm{L}_1, \mathrm{R}_1, \mathrm{U}_1, \mathrm{D}_1, \mathrm{D}_2, \mathrm{U}_2)$, vehicle $\mathrm{L}_1$ with the highest priority can directly cross the CA, while vehicle $\mathrm{R}_1$ needs to give way to vehicle $\mathrm{L}_1$ since potential merge collision may occur. Now that we focus on vehicle $\mathrm{D}_1$, it may conflict with both vehicle $\mathrm{R}_1$ and vehicle $\mathrm{U}_1$. In this case, $\mathrm{R}_1$ and $\mathrm{U}_1$ first plan their own trajectories and then update occupation time of all collision regions along their paths. We use a collision time table $\mathcal{T}_\mathrm{time}$ which can be stored in the coordinator or shared among vehicles to update occupation times. For example, vehicle $\mathrm{R}_1$ predicts that it will occupy collision region $\mathrm{C}_\mathrm{RD, DL} = [s_\mathrm{RD, DL}^\mathrm{low}, s_\mathrm{RD, DL}^\mathrm{high}]$ during time duration $[t_\mathrm{RD, DL}^\mathrm{left}, t_\mathrm{RD, DL}^\mathrm{right}]$, then the collision time table can be updated as $\mathcal{T}_\mathrm{time}[\mathrm{DL}][\mathrm{RD}] = [t_\mathrm{RD, DL}^\mathrm{left}, t_\mathrm{RD, DL}^\mathrm{right}]$. Similarly, vehicle $\mathrm{U}_1$ will occupy collision region $\mathrm{C}_\mathrm{UD, DL} = [s_\mathrm{UD, DL}^\mathrm{low}, s_\mathrm{UD, DL}^\mathrm{high}]$ during time duration $[t_\mathrm{UD, DL}^\mathrm{left}, t_\mathrm{UD, DL}^\mathrm{right}]$, so $\mathcal{T}_\mathrm{time}[\mathrm{DL}][\mathrm{UD}]$ can be set as $[t_\mathrm{UD, DL}^\mathrm{left}, t_\mathrm{UD, DL}^\mathrm{right}]$.

All the blocked regions along the path can be marked in the S-T graph, as shown in Fig.~\ref{fig:stplanner}. The red and green areas represent the time-spatial resources occupied by high-priority vehicles $\mathrm{R}_1$ and $\mathrm{U}_1$, respectively. Thus, the local trajectory planning is equivalent to finding a speed profile that does not intersect with these blocked areas. The solid line means vehicle $\mathrm{D}_1$ gives way to vehicle $\mathrm{U}_1$ and enters the collision region $\mathrm{C}_\mathrm{DL, UD}$ right after vehicle $\mathrm{U}_1$ exits. It should be noted that though in this case vehicle $\mathrm{D}_1$ could preempt the collision region $\mathrm{C}_\mathrm{DL,UD}$ (the dashed line), this is unacceptable since vehicles must follow the given priority sequence in order to avoid possible deadlocks. This is also a clear indication that priority scheduling is essential for coordination efficiency. 


Since ego vehicle must give way to high-priority vehicles, the speed profile of ego vehicle must below the blocked areas to avoid lateral collisions. Denote $\mathcal{C_\mathrm{ts}}$ the set of coordinates of the lower right corner of occupied resource blocks in the S-T graph, i.e.,
\begin{align}
	\mathcal{C_\mathrm{ts}} = \big\{(t_\mathrm{right}, s_\mathrm{low})\big| &(t_\mathrm{left}, t_\mathrm{right}) \in \mathcal{T}_\mathrm{times}[p_\mathrm{ego}], \notag\\&(s_\mathrm{low}, s_\mathrm{high}) \in \mathcal{T}_\mathrm{collision}[p_\mathrm{ego}]  \big\}
\end{align}
the lateral safety constraints can be expressed using the following linear inequalities:
\begin{align}
	s(t_\mathrm{right}) \leq s_\mathrm{low}, \forall (t_\mathrm{right}, s_\mathrm{low}) \in \mathcal{C_\mathrm{ts}} \label{eq:mctslateral}
\end{align}

In general, the local trajectory planning problem of each vehicle $i \in \mathcal{N}$ can be summarized as follows,
\begin{align*}
	\mathscr{P}_\mathrm{distributed}^i: \, \, &\min_{u_i(t)} \,\, \sum_{i=1}^{N} \sum_{t=0}^{T-1} (v_i(t) - v_\mathrm{max})^2 \\
	\mathrm{s.t.} &\, \, \,  \eqref{eq:kinematic_1} - \eqref{eq:kinematic_4},\,\eqref{eq:csrear_end},\,\eqref{eq:mctslateral}
\end{align*}
This problem can be cast as a low-complexity Quadratic Programming (QP) problem with only a few linear collision-free constraints. In this way, the centralized MIQP problem $\mathscr{P}_\mathrm{centralized-OCP}$ is decomposed into a sequence of distributed QP problems, i.e., $\{\mathscr{P}_\mathrm{distributed}^i | i \in \mathcal{N}\}$. 

To further improve the overall traffic efficiency, in the next section, we present the high-level optimization problem in which we seek the optimal priority sequence (i.e., the order to solve the above sub-problems).

\section{High-level Centralized Priority Scheduler}\label{sec:high_level_planner}
\subsection{Simplified Speed Profiles}
The high-level planner aims at finding an optimal priority sequence from a systematic point of view.
The priority assignment problem can be mathematically written as the following combinatorial optimization problem:
\begin{align*}
	\mathscr{P}_\mathrm{priority}: \, \, &\min_{\mathbf{o} \in \mathcal{O}} \,\, t_\mathrm{leave}(\mathbf{o})
\end{align*}
where $t_\mathrm{leave}(\mathbf{o})$ denotes the total passing time the vehicles follow a given priority sequence $\mathbf{o}$, $\mathcal{O}$ is the set of all possible priority sequences. However, the calculation of $t_\mathrm{leave}(o)$ is time-consuming. Specifically, for a given priority sequence $o$, $t_\mathrm{leave}(o)$ can only be obtained after all vehicles have planned their trajectories. This direct evaluation approach takes over-long problem-solving time in real-world implementations. For example, if a vehicle needs $\tau_c$ seconds to plan its local trajectory, the evaluation of a given priority sequence of $N$ vehicles consumes $N\tau_c$ seconds. To find the optimal priority sequence, we need to traverse all $N!$ solutions, which requires about $N!N\tau_c$ seconds. To address this issue, we present an efficient evaluation method, as shown in Fig.~\ref{fig:simplifiedstplanner}.

The speed profiles in Fig.~\ref{fig:stplanner} are approximated by straight lines, i.e.,
\begin{align}
	s_i(t) = \begin{cases}
		s_0^i,             & t_0 \leq t \leq t_\mathrm{acc}^i \\
		v_\mathrm{max}(t - t_\mathrm{acc}^i) + s_0^i,   & t_\mathrm{acc}^i \leq t \leq t_\mathrm{end}^i
	\end{cases}
\end{align}
where $t_\mathrm{acc}^i$ and $t_\mathrm{end}^i$ are the starting time and existing time, respectively. The straight line with slop $v_\mathrm{max}$ can be denoted as $s = v_\mathrm{max} t + b$ and must below all the blocked regions, such that
\begin{align}
	b_i = \min \{s - v_\mathrm{max}t | (t, s) \in \mathcal{C}_\mathrm{ts}\}  \label{eq:bi}
\end{align}
To avoid rear-end collisions with preceding vehicle on the same lane, we have
\begin{align}
	b_i^* = \min \{b_i, b_\mathrm{pre}^* - L_\mathrm{safe}\}
\end{align}
where $b_\mathrm{pre}^*$ is the optimal parameter of preceding vehicle, $L_\mathrm{safe}$ is the safety distance. Hence, $t_\mathrm{acc}^i$ and $t_\mathrm{end}^i$ can be calculated as follows,
\begin{align}
	t_\mathrm{acc}^i = \frac{s_i^0 - b_i^*}{v_\mathrm{max}}, \, \, \, t_\mathrm{end}^i = \frac{s^\mathrm{end}_i - b_i^*}{v_\mathrm{max}} 
\end{align}
Also, the occupation time of the collision regions along its path can be calculated in the same way. For a given priority sequence $\mathbf{o}$, $t_\mathrm{leave}(\mathbf{o})$ can be calculated as:
\begin{align}
	t_\mathrm{leave}(\mathbf{o}) = \max_{i \in \mathcal{N}} \{t_\mathrm{end}^i \, |\, \mathbf{o}\} \label{eq:t_leave}
\end{align}
Though the simplified speed profiles do not satisfy kinematic constraints \eqref{eq:kinematic_1} - \eqref{eq:kinematic_2}, they can evaluate the quality of a priority sequence simply using simple calculations.

\begin{figure}[htbp]
	\centering
	\includegraphics[width=0.8\linewidth]{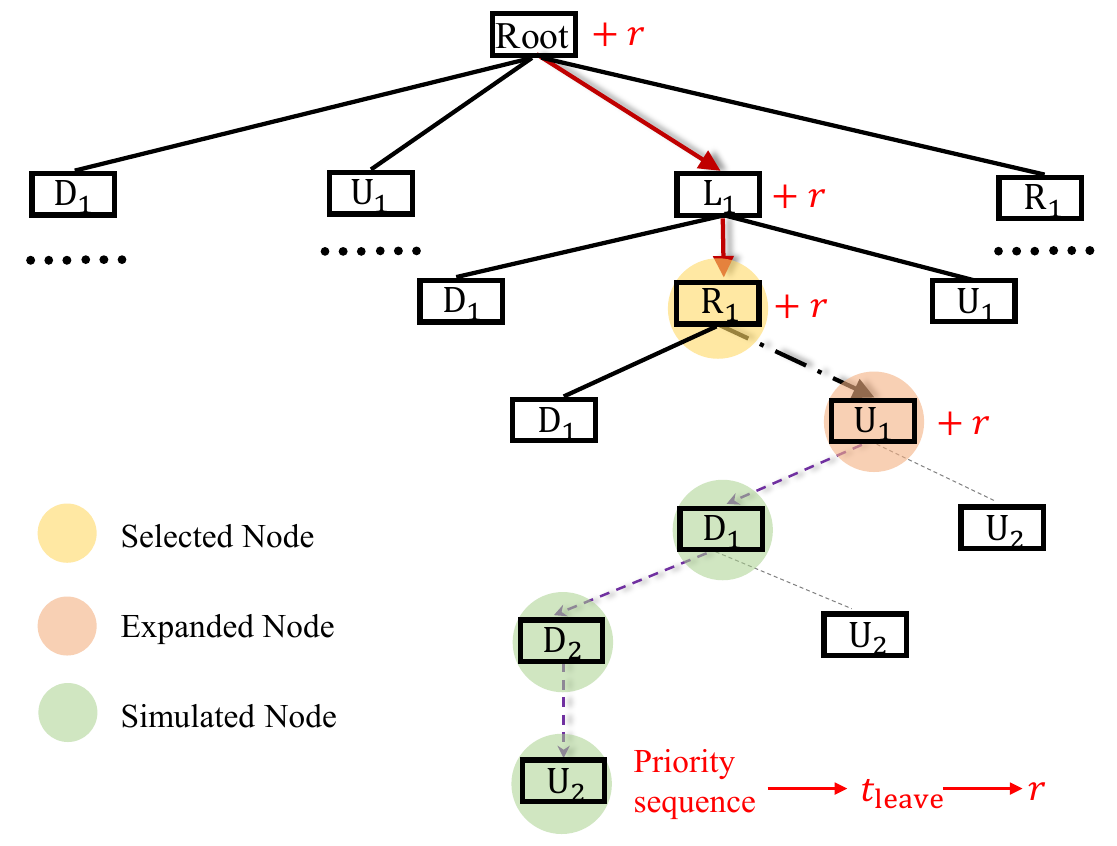}
	\caption{Construction of the MCTS based solution tree.}
	\label{fig:mctsillustration}
\end{figure}

\subsection{Monte Carlo Tree Search Based Priority Scheduler}
Problem $\mathscr{P}_\mathrm{priority}$ is still challenging and it is typically impossible to traverse all priority sequences within the limited computational budget, especially when a large number of vehicles within the coordination area. In the following, we present an MCTS based approach which could efficiently find high-quality sequences within a short planning time. The effectiveness of MCTS has been shown in many sequential decision-making problems such as games, planning, and control\cite{mcts,alphago,mcts_driving}.

The key idea of MCTS is to iteratively build a search tree by estimating the values of tree nodes through random simulations \cite{mcts}. The computational budget (e.g., planning time constraint or iteration constraint) is pre-defined, such that MCTS can always be stopped and return a solution, though might not be optimal. Four steps are applied per search iteration: selection, expansion, simulation, and backpropagation, as shown in Fig.~\ref{fig:mctsillustration}. Firstly, select the most valuable and expandable node, i.e., vehicle $\mathrm{R}_1$, according to a certain \textit{tree search policy}. Secondly, expand a new node that can be selected in the next round, i.e., $\mathrm{U}_1$. Thirdly, simulate by randomly sampling a feasible priority sequence, e.g., $(\mathrm{L}_1, \mathrm{R}_1, \mathrm{U}_1, \mathrm{D}_1, \mathrm{D}_2,\mathrm{U}_2)$ shown in Fig.~\ref{fig:systemmodel}. Calculate the cost $t_\mathrm{leave}$ according to \eqref{eq:bi} - \eqref{eq:t_leave} and convert it into reward $r$ using a certain \textit{reward function}. Then, backpropagate the current reward $r$ to the selected node $\mathrm{U}_1$ and its ancestors $(\mathrm{root}, \mathrm{L}_1, \mathrm{R}_1)$. 

In this paper, we adopt the commonly used tree search policy called Upper Confidence Bounds for Trees (UCT), which can be expressed as:
\begin{align}
	c^* = \max_{c \in \mathcal{A}_p} \, \frac{R_c}{n_c} + \epsilon \sqrt{\frac{\ln n_p}{n_c}} \label{eq:UCT}
\end{align}
where $p$ is a parent node, $\mathcal{A}_p$ is the set of possible children (the first uncoordinated vehicle on each lane), and $c$ is a child of parent node $p$. $R_c$, $n_c$, $n_p$ represent the cumulative rewards of child $c$, the number of visits of child $r$, and the number of visits of parent $p$, respectively. The first term of \eqref{eq:UCT} is called the exploitation term, which means a child with higher cumulative rewards is more likely to be selected. While the second term is called the exploration term, which encourages agent to select less visited nodes that may result in better results. The constant $\epsilon > 0$ makes a trade-off between exploitation and exploration. 

The reward function is essential for the MCTS algorithm. Generally, the first term of \eqref{eq:UCT} should be within $[0, 1]$. This can be easily guaranteed in problems such as games since the reward is set as $\{\mathrm{win=1}, \mathrm{draw=0.5}, \mathrm{loss=0}\}$. In our priority scheduling problem, we normalize the reward of each node $c$ whose parent is node $p$ to $[0, 1]$ with the following formulation,
\begin{align}
	r_p^c = \frac{f_p^c - f_{p,\mathrm{min}}}{f_{p,\mathrm{max}} - f_{p,\mathrm{min}}} \label{eq:reward}
\end{align}
where $f_p^c=- t_\mathrm{leave}^c$ is the instant reward obtained after each simulation step, $f_{p,\mathrm{min}}$ and $f_{p,\mathrm{max}}$ are the minimal and maximal reward among all children nodes of node $p$ respectively. Note that $f_{p, \mathrm{max}}$ and $f_{p, \mathrm{min}}$ are dynamically determined as the MCTS algorithm progresses. Specifically, at the current simulation step, if the parent node $p$ has no child nodes, then we set $f_{p,\mathrm{min}} = f_{p,\mathrm{max}} = f_\mathrm{sim}$, where $f_\mathrm{sim}$ is a temporal reward obtained by randomly sampling a feasible sequence from parent node $p$. On the other hand, if $p$ already has existing children, then $f_{p,\mathrm{min}}$ and $f_{p,\mathrm{max}}$ are set to be the minimal and maximal reward among these children. The proposed MCTS-based priority scheduler is summarized in Algorithm \ref{alg:mcts}. 

\begin{algorithm}[h]
	\caption{MCTS Based Priority Scheduler.}
	\label{alg:mcts}
	\KwIn{$\mathcal{N}$, $\mathcal{T}_\mathrm{collision}$, $\mathcal{T}_\mathrm{time}$}
	\KwOut{priority sequence $\mathbf{o}^*$}
	\For{Pre-defined computational budget}{
		\textit{node} $\leftarrow$ \textit{root}, $c^*$ = $\emptyset$\;
		\While{\textrm{node} is not terminal}{
			\If{node is expandable}{
				Expand a child $c$ for \textit{node}\;
				$c^*$ $\leftarrow$ $c$, break loop\;}
			\Else{Find a best child $c$ acoording to \eqref{eq:UCT}\;
				\textit{node} $\leftarrow$ $c$, continue\;
			}
		}
		Run a simulation from $c^*$\;
		Calculate reward according to \eqref{eq:bi} - \eqref{eq:t_leave} and  \eqref{eq:reward}\;
		Backpropogate rewards of current branch\;
	}
	Set $\epsilon = 0$. Starting at the root node, recursively apply \eqref{eq:UCT} to descend through the tree until finding an optimal priority sequence $\mathbf{o}^*$\;
\end{algorithm}

\begin{figure*}[t]
	\centering
	\includegraphics[width=1\textwidth]{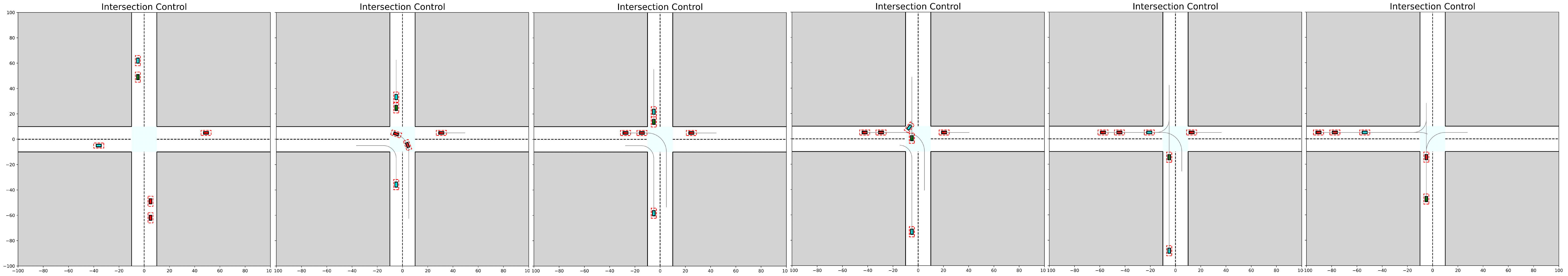}
	\caption{A capture of coordination animation under CS strategy ($t_\mathrm{leave} = 18.1$s).}
	\label{fig:csvideo}
\end{figure*}

\begin{figure*}[t]
	\centering
	\includegraphics[width=1\textwidth]{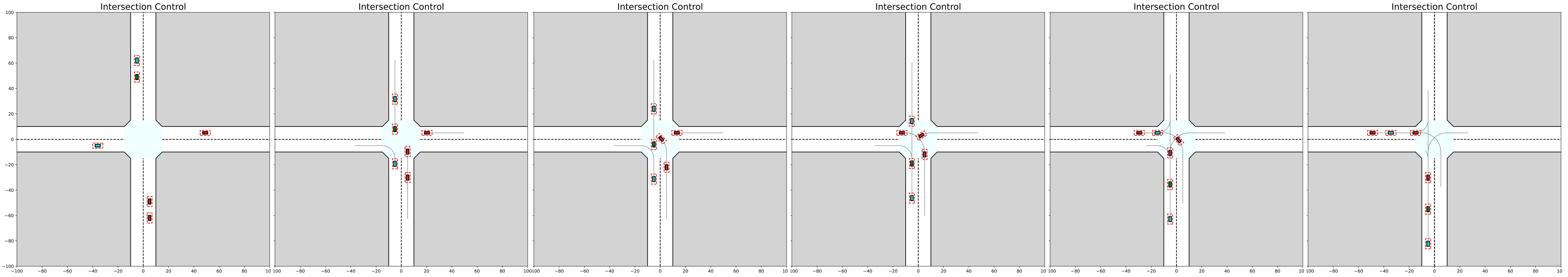}
	\caption{A capture of coordination animation under the proposed bi-level strategy ($t_\mathrm{leave} = 14.9$s).}
	\label{fig:mctsvideo}
\end{figure*}

\begin{figure*}[t]
	\centering
	\begin{minipage}[t]{0.48\textwidth}
		\centering
		\includegraphics[width=0.95\linewidth]{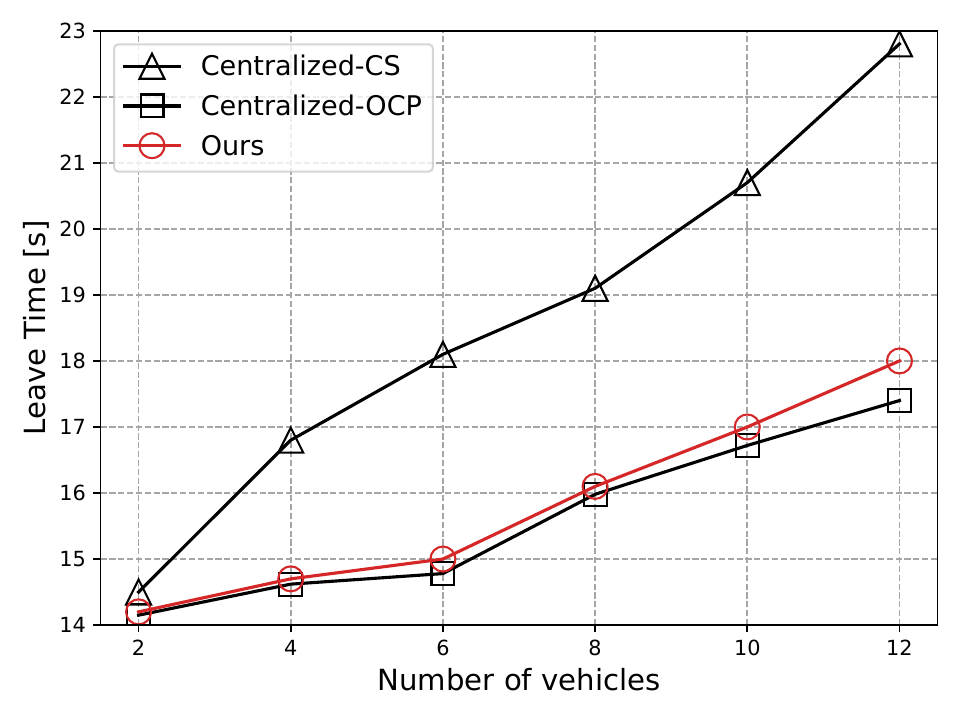}
		\caption{Comparison of coordination efficiency.}
		\label{fig:leavetime}
	\end{minipage}
	\hfill
	\begin{minipage}[t]{0.48\textwidth}
		\centering
		\includegraphics[width=0.95\linewidth]{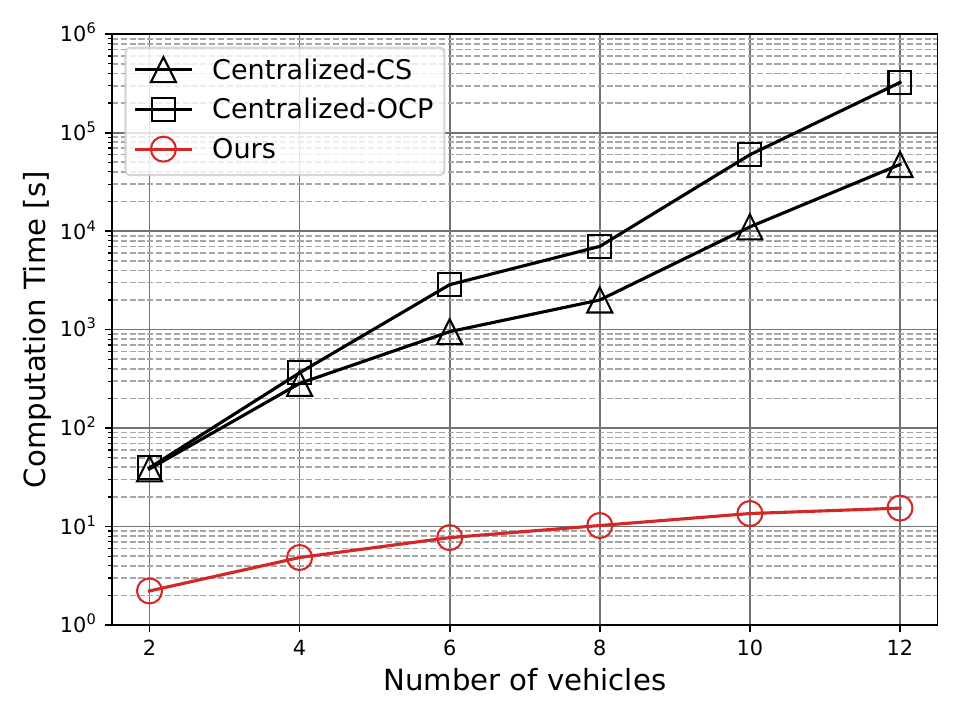}
		\caption{Comparison of computational complexity.}
		\label{fig:computationtime}
	\end{minipage}
\end{figure*}

\section{Simulation Results and Experiments}\label{sec:results}
\subsection{Simulation Settings}
In this section, we conduct simulations to evaluate the efficiency and adaptability of the proposed bi-level coordination strategy. To compare the coordination efficiency in sub-static scenarios, both the centralized-CS and centralized-OCP strategies described in Section \ref{sec:centralizedcs} are considered as benchmarks. In addition, we implement traditional traffic signal control and a FIFO-based distributed strategy to validate traffic throughput in continuous traffic flows with different arrival rates. In traffic signal control, signal timing is predetermined using historical traffic data, and only non-conflicting vehicles are allowed to enter the CA during each green light phase.

All experiments are conducted with Python 3.8 on a personal computer with Intel i7-10700F 2.9 GHz CPU and 16 GB of RAM. We construct a signalized intersection using the Simulation of Urban MObility (SUMO) platform and use Python to interact with SUMO via the TraCI library. The centralized-CS problem $\mathscr{P}_\mathrm{centralized}$, the centralized-OCP problem $\mathscr{P}_\mathrm{centralized-OCP}$, and the distributed trajectory planning problem $\{\mathscr{P}_\mathrm{distributed}^i|i\in \mathcal{N}\}$ are solved using the MOSEK solver. We also use Python API to access the MOSEK optimizer. The chosen parameters for the vehicles and the road structure are shown in Table \ref{tab:table_1}. Videos of our simulation results can be found at the supplementary materials, or online at \url{https://youtu.be/WYAKFMNnQfs}.

\begin{table}[h]
	\centering
	\caption{Simulation Parameters}
	\label{tab:table_1}
	\begin{tabular}{cc}
		\toprule
		\textbf{Parameters}               &    Value   \\
		\midrule
		radius of the coordination area   &    100 m   \\
		lane width                        &     10 m    \\
		left-turning radius               &     15 m    \\
		right-turning radius              &      10 m  \\
		total simulation time $T_\mathrm{total}$ & 40 s \\
		slot duration $\Delta t$           &       0.1 s        \\
		maximum velocity $v_\mathrm{max}$   &     15 m/s  \\
		maximum acceleration $u_\mathrm{max}$&     5 $\mathrm{m/}\mathrm{s}^2$ \\
		minimum acceleration $u_\mathrm{min}$&    - 5 $\mathrm{m/}\mathrm{s}^2$ \\
		vehicle geometry                   &     2 m $\times$ 4 m   \\
		safety box                         &     4 m $\times$ 8 m   \\
		safety distance $L_\mathrm{safe}$&    8 m   \\
		big-M parameter                       &    10000 \\
		UCT factor $\epsilon$              &      $\sqrt{2}$  \\
		MCTS iteration constraint          &      10000 \\
		\bottomrule
	\end{tabular}
\end{table}

\subsection{Comparison of Different Coordination Strategies}
We first consider the coordination scenario illustrated in Fig.~\ref{fig:systemmodel} for comparison of the centralized-CS and our bi-level strategies. Fig.~\ref{fig:csvideo} shows a capture of coordination animation under the centralized-CS strategy. We observe that vehicle $\mathrm{L}_1$ crosses the coordination area first, followed by high-priority vehicles $\mathrm{D}_1$ and $\mathrm{D}_2$, causing the remaining vehicles to decelerate or stop before the entry point. Once $\mathrm{D}_1$ and $\mathrm{D}_2$ exit the coordination area, vehicles $\mathrm{U}_1$ and $\mathrm{U}_2$ enter, with vehicle $\mathrm{R}_1$ being the last to cross. The resulting priority sequence is $(\mathrm{L}_1, \mathrm{D}_1, \mathrm{D}_2, \mathrm{U}_1, \mathrm{U}_2, \mathrm{R}_1)$, and the optimal total travel time (i.e., the time when the last vehicle leaves the coordination area) is $t_\mathrm{leave} = 18.1$s.

Fig.~\ref{fig:mctsvideo} shows an animation of the coordination process under the proposed bi-level strategy. The priority sequence obtained by the high-level planner is $(\mathrm{L}_1, \mathrm{U}_1, \mathrm{D}_1, \mathrm{R}_1, \mathrm{U}_2, \mathrm{D}_1)$. The vehicles then plan their trajectories sequentially based on this priority configuration. The collision set can accommodate up to 4 vehicles simultaneously, suggesting that the proposed collision region model can fully utilize time-spatial resources. The resulting total travel time is $t_\mathrm{leave} = 14.9$s, which represents a 17.7\% increase in traffic efficiency compared to the centralized-CS strategy. We also evaluate the priority sequence $(\mathrm{L}_1, \mathrm{R}_1, \mathrm{U}_1, \mathrm{D}_1, \mathrm{D}_2, \mathrm{U}_2)$ shown in Fig.~\ref{fig:systemmodel}, which results in a total travel time of approximately 16.7s. This highlights the significance of assigning priorities for coordination efficiency.

Fig.~\ref{fig:leavetime} demonstrates the coordination efficiency of these three strategies with respect to the number of vehicles. We observe that our strategy significantly outperforms the centralized-CS strategy, as the collision model we propose allows vehicles to share the CA simultaneously. As the number of vehicles increases, the efficiency of the centralized-CS strategy decreases because more vehicles have to yield to conflicting vehicles. The centralized-OCP strategy, which integrates our collision region model, can always achieve optimal coordination performance. Notably, our bi-level strategy achieves near-optimal coordination performance, with a small gap observed between our approach and the centralized-OCP strategy. This gap may be attributed to (i) sub-optimal priority sequences caused by MCTS and (ii) sub-optimal trajectories caused by sequential optimization.

Fig.~\ref{fig:computationtime} shows the maximum computation time with respect to the number of vehicles for these strategies. Our results show that when solving a local trajectory planning problem (i.e., $\mathscr{P}_\mathrm{distributed}^i, i\in\mathcal{N}$) over the total planing horizon $T_\mathrm{total} = 40s$ (i.e., $T_\mathrm{total}/\Delta t = 400$ slots), the MOSEK solver typically takes about 0-1.5s, and the computation time increases linearly with the number of vehicles. It is worth noting that in real-world deployments, receding horizon control methods (i.e., model predictive control) can be employed in our framework to solve the coordination problem at each time slot over a short planning horizon, thus reducing computation time. Additionally, the MCTS-based high-level planner always returns a priority sequence within 0.1s. In contrast, the centralized strategies suffer from high computational complexity, with computation time increasing exponentially with the number of vehicles. This is because $\mathscr{P}_\mathrm{centralized-CS}$ and $\mathscr{P}_\mathrm{centralized-OCP}$ are MIQP problems, which are computationally prohibitive in large problem sizes. For instance, when dealing with 12 vehicles, the centralized-CS strategy takes approximately 13.2 hours to find a solution, while our approach achieves near-optimal results in only 14.1 seconds.

\subsection{Evaluation of Throughput in Continuous Traffic Flow}
\begin{figure}[h]
	\centering
	\includegraphics[width=\linewidth]{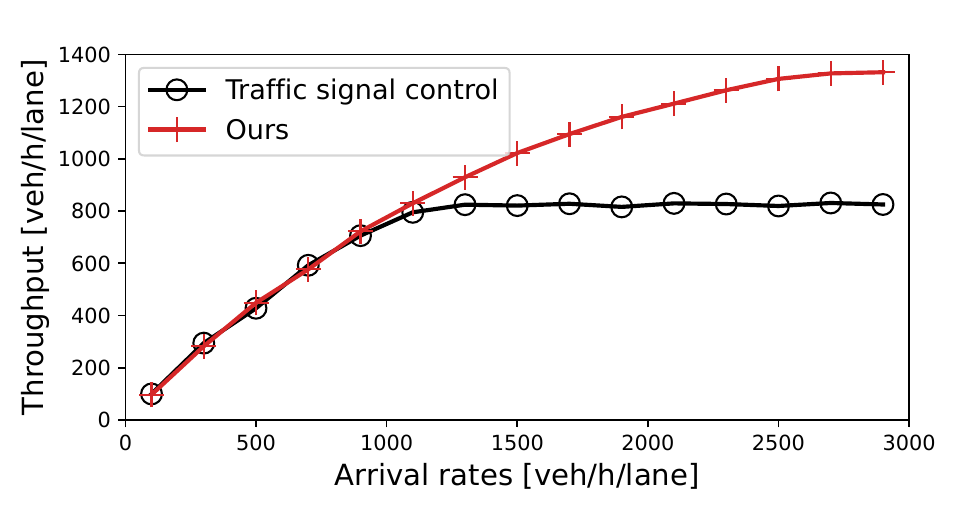}
	\caption{Traffic throughput without left-turning vehicles.}
	\label{fig:throughputprob0}
\end{figure}

The proposed bi-level coordination strategy is also applied to continuous traffic flow. The vehicles arrival is assumed to be a Poisson process. In Fig.~\ref{fig:throughputprob0}, we vary the arrival rates from 100 to 3000 vehicles/h/lane to compare the traffic throughput of the traffic signal control and our strategy under different traffic loads. For each arrival rate, the traffic throughput is averaged over the number of coordinated vehicles of 100 independent trials, and each trial simulates a 1-hour traffic process. To present the upper bound of throughput under a certain strategy, Fig.~\ref{fig:throughputprob0} does not consider left-turning vehicles. It can be seen that, in low traffic loads (100 to 800 veh/h/lane), both traffic signal control and the proposed strategy can provide enough coordination capacity for incoming traffic volumes. However, the traffic signal control soon reaches its capacity (about 800 veh/h/lane) when the arrival rate is 1200 veh/h/lane. In contrast, the proposed strategy reaches its limit when the arrival rate is about 2800 veh/h/lane and can provide a much higher serving rate of up to 1400 veh/h/lane. 

\begin{figure}[h]
	\centering
	\includegraphics[width=0.95\linewidth]{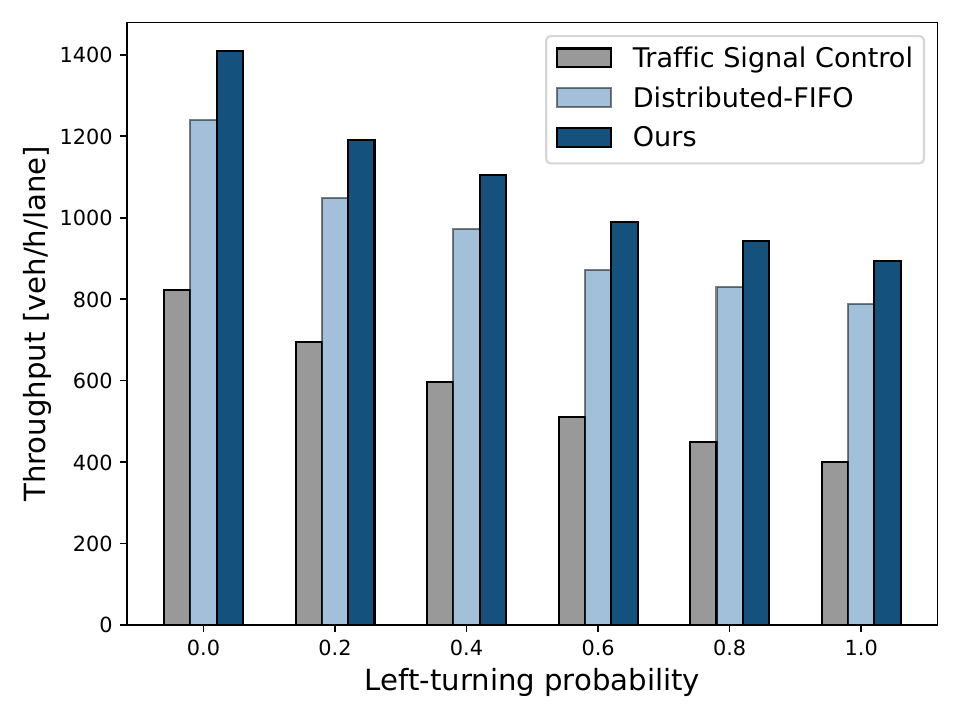}
	\caption{Traffic throughput with different left-turning probabilities.}
	\label{fig:throughputdifferentprobs}
\end{figure}

Fig.~\ref{fig:throughputdifferentprobs} further examines the influence of different left-turning probabilities on traffic throughput. For each traffic pattern, we simulate a 1-hour traffic process with a arrival rate of 3000 veh/h/lane. It is evident that compared to traffic signal control, our strategy can significantly improve traffic throughput, regardless of the traffic patterns. We observe that an increase in left-turning vehicles results in a decline in throughput. This is because left-turning paths occupy more collision regions and intersect with many paths, leading to more vehicles giving way to left-turning ones. Furthermore, we compare the efficiency of our MCTS-based priority scheduler with the FIFO method. For fairness, the distributed-FIFO strategy uses the same trajectory planner as ours. We observe that our MCTS-based scheduler has an approximately 12\% improvement in traffic throughput compared to the FIFO method.

\section{Conclusion}\label{sec:conclusion}
This work presented a priority-based bi-level coordination framework for intersection control of CAVs. Two main issues to be addressed were traffic throughput and computational complexity. To reduce computation time, we decomposed the optimal coordination problem into the priority scheduling and trajectory planning. An MCTS based priority scheduler was proposed to efficiently find a near-optimal priority sequence. Based on the given priority configuration, a low-complexity distributed trajectory planning problem with the aim of minimizing travel time was also formulated. Simulation results have shown that the proposed strategy achieves comparable coordination performance as the centralized-OCP strategy and exhibits good scalability. 

Future research directions involve an extension of the presented framework to accommodate system uncertainties. We plan to explore robust controllers based on model predictive control (MPC), control barrier function (CBF), and covariance steering (CS). Moreover, we will take into account realistic coordination scenarios and incorporate non-cooperative traffic participants. Addressing these challenges will make the priority scheduling and distributed trajectory planning problems more interesting but also more challenging to solve.

\bibliographystyle{IEEEtran}
\bibliography{ref}

\end{document}